\documentclass[sigconf,screen,nonacm]{acmart}
\acmConference[ICSE 2024]{46th International Conference on Software Engineering}{April 2024}{Lisbon, Portugal}
\AtBeginDocument{%
  }

\usepackage{enumitem}
\usepackage{listings}
\usepackage{tikz}
\usetikzlibrary{shapes.geometric, arrows, positioning, fit, shapes.misc}
\usepackage{xspace}
\usepackage{cleveref}
\usepackage{multirow}
\usepackage{makecell}
\usepackage{lipsum}
\usepackage{subcaption}
\usepackage{amsmath}
\usepackage{xcolor}
\usepackage{balance}
\usepackage{tcolorbox}
\usepackage{tabularx,booktabs}
\usepackage[normalem]{ulem}
\def\product{\textsc{SAP HANA}\xspace}
\def\pwork{Bach et al.~\cite{Bach2022}\xspace}

\newcolumntype{Y}{>{\centering\arraybackslash}X}

\makeatletter
\DeclareDocumentCommand\highLightSwitch{m m}{%
  \setlength{\fboxsep}{1pt}\color{#1}\colorbox{#2}
}
\makeatother

\setcopyright{acmcopyright}
\copyrightyear{2018}
\acmYear{2018}
\acmDOI{XXXXXXX.XXXXXXX}

\acmPrice{15.00}
\acmISBN{978-1-4503-XXXX-X/18/06}

\setcopyright{none} 
\begin{document}

\title{Just-in-Time Flaky Test Detection via Abstracted Failure Symptom Matching}


\settopmatter{printacmref=false, printccs=false, printfolios=false} 

\author{Gabin An}
\affiliation{%
  \institution{KAIST}
  \city{Daejeon}
  \country{Republic of Korea}}
\email{agb94@kaist.ac.kr}

\author{Juyeon Yoon}
\affiliation{%
  \institution{KAIST}
  \city{Daejeon}
  \country{Republic of Korea}}
\email{juyeon.yoon@kaist.ac.kr}

\author{Thomas Bach}
\affiliation{%
  \institution{SAP}
  \city{Walldorf}
  \country{Germany}}
\email{thomas.bach03@sap.com}

\author{Jingun Hong}
\affiliation{%
  \institution{SAP Labs Korea}
  \city{Seoul}
  \country{Republic of Korea}}
\email{jingun.hong@sap.com}

\author{Shin Yoo}
\affiliation{%
  \institution{KAIST}
  \city{Daejeon}
  \country{Republic of Korea}}
\email{shin.yoo@kaist.ac.kr}


\begin{abstract}
We report our experience of using failure symptoms, such as error messages or 
stack traces, to identify flaky test failures in a Continuous Integration (CI) 
pipeline for a large industrial software system, \product. Although failure 
symptoms are commonly used to identify similar failures, they have not 
previously been employed to detect flaky test failures. Our hypothesis is that 
flaky failures will exhibit symptoms distinct from those of non-flaky failures. 
Consequently, we can identify recurring flaky failures, without
rerunning the tests, by matching the failure symptoms to those 
of historical flaky runs. This can significantly reduce the need for test reruns, ultimately 
resulting in faster delivery of test results to developers. To facilitate the 
process of matching flaky failures across different execution instances, we 
abstract newer test failure symptoms before matching them to the known patterns 
of flaky failures, inspired by previous research in the 
fields of failure deduplication and log analysis. We evaluate our symptom-based 
flakiness detection method using actual failure symptoms gathered from 
CI data of \product during a six-month period. Our method shows the 
potential of using failure symptoms to identify recurring flaky failures, achieving a precision of at least 96\%, while saving approximately 
62\% of the machine time compared to the traditional rerun strategy. Analysis 
of the false positives and the feedback from developers underscore the 
importance of having descriptive and informative failure 
symptoms for both the effective deployment of this symptom-based approach and 
the debugging of flaky tests. 
\end{abstract}





\maketitle

\section{Introduction}



\product is an in-memory Database Management System (DBMS) that is used by many of the largest enterprises around the world, offering both an on-premise installation and a database-as-a-service solution. A commercial database service like \product requires tremendous effort in testing because the cost incurred by production errors is prohibitively expensive. Therefore, \product is tested both intensively and systematically against every incoming code change to detect bugs as early as possible in its Continuous Integration (CI) environment~\cite{Bach2022}.

As pointed out in previous work~\cite{Bach2022}, one of the main challenges of testing \product is \emph{flaky tests}~\cite{zheng_research_2021, parry_survey_2021} that both fail and pass against the same version of the source code, making their results less actionable. The flakiness of tests not only diminishes the reliability of test results~\cite{parry_surveying_2022}, but also increases the cost of testing in \product. This is because failed test cases are often re-executed multiple times to determine if they consistently fail or not, in the so-called \emph{rerun strategy}~\cite{google_flaky,spotify_flaky}. Our investigation on the pre-submit testing data of \product shows that 87\% of test failures are discovered to be flaky, and a significant portion of the total testing time per test run, with a maximum of 67\% and an average of 10\%, is spent on just rerunning the failed tests.
Such increase in testing time means that developers have to wait longer for the test results, which in turn reduces their productivity~\cite{bell_deflaker_2018,parry_surveying_2022}.

To address these challenges, researchers have proposed various flaky test detection techniques that do not require reruns. Some techniques rely on only static features, such as the vocabulary of source code~\cite{pinto_what_2020, camara_what_2021, haben_replication_2021} or test smells~\cite{pontillo_static_2022}. While these approaches are useful for identifying source code patterns that are indicative of potential flakiness in a test suite, they cannot accurately handle a single test case that can result in both flaky and non-flaky failures. Note that \emph{flaky failure} refers to a failure that is not consistently reproduced against the same version of the program.
Dynamic techniques, on the other hand, focus on failing test executions. DeFlaker~\cite{bell_deflaker_2018}, for example, determines the flakiness of test failure based on whether it executes the recently changed code or not. However, by design, this technique is unable to detect the flaky failures whose coverage overlaps with the changed code. Moreover, it requires precise code coverage, the cost of which makes it less practical for use in extensive testing of large-scale projects.

Meanwhile, there has been a significant amount of research on failure deduplication~\cite{dang2012,lerch2013finding,bartz2008finding, Brodie, Brodie2005, Modani2007}, which uses easily obtainable failure symptoms such as stack traces or error messages to identify and group failures that have the same root cause. These studies have demonstrated that failure symptoms can be valuable information sources for identifying duplicate failures. In the context of flaky tests, it is also assumed that failure symptoms contain information about the cause of flakiness. For example, Flakes, a flaky test management system offered by CloudBuild~\cite{Esfahani2016}, uses failure symptoms in the bug report assignment process, linking multiple flaky tests with similar error messages to the same bug report~\cite{lam_study_2020}. Additionally, FlakeRepro~\cite{Leesatapornwongsa2022}, a flaky test reproduction technique, uses error messages to determine whether a flaky failure has been successfully reproduced. However, to the best of our knowledge, failure symptoms have not yet been explicitly used to detect recurring flaky failures during the testing process in CI systems, despite their usefulness in failure deduplication.

In this paper, we propose a lightweight and black-box approach to detect recurring flaky tests in a CI environment using failure symptoms and information from historical test executions. Our approach is based on the idea that failure symptoms can be linked to the root cause of the flakiness and therefore can be used as an indicator of the flakiness. During the CI cycle, we gather the symptoms of the flaky failures, which are discovered by the rerun strategy, and subsequently use them to predict whether a new failure is flaky or not. If the symptoms of a new failure have been frequently observed in previous flaky failures, the new failure is regarded as flaky without re-executions of tests. To increase the chances of matching symptoms from failures with the same root cause across different execution instances, we abstract the symptoms to include only the most relevant information related to the failure.
When evaluated with historical CI data of \product, our approach achieves a 96\% precision and a 76\% recall in detecting flaky failures. The ablation study outcomes reveal that the abstraction of symptoms has a great impact on the performance, increasing recall from 50\% to 76\% while retaining a similar level of precision. Additionally, we find that our prediction can potentially save about 62\% of machine time spent for rerunning the failed tests.
Furthermore, the abstraction also enables the grouping of similar and recurring flaky failures, which can help the manual investigation by developers. Overall, these results demonstrate the substantial promise of using failure symptoms to identify recurring flaky failures in a CI pipeline.


We summarise the contributions of this work as follows:

\begin{itemize}[leftmargin=*]
\item \textbf{Novel Black-box Flakiness Detection}: While failure symptoms have previously been used to group similar failures, we are the first, to the best of our knowledge, to apply and evaluate their effectiveness specifically in the context of flakiness detection.
\item \textbf{Enhanced Detection Through Abstraction}: We demonstrate the benefits of using two abstraction methods, number masking from the log analysis domain and stack trace purification from the failure deduplication domain, in detecting flaky failures.
\item \textbf{Real-World Evaluation}: We extensively evaluate our approach using a substantial volume of real-world failure data obtained from \product. 
\end{itemize}

The remainder of this paper is organised as follows. Section~\ref{sec:background} provides background information on our target software, \product, and discusses the problem of flaky tests. Section~\ref{sec:method} describes our method for identifying flaky failures using failure symptoms. We describe the evaluation settings in Section~\ref{sec:setup}, and present the results and discussions in Section~\ref{sec:results} and Section~\ref{sec:discussion}, respectively. In Section~\ref{sec:rw}, we survey related work in the areas of flaky test detection and failure deduplication. Finally, we conclude in Section~\ref{sec:conclusion}.

\section{Background}
\label{sec:background}

This section outlines the testing pipeline, and discusses the issue of flaky tests, in our target project, \product, a large-scale DBMS that consists of millions of lines of C++ code and about 1 million test cases~\cite{Bach2022}.

\subsection{Testing Pipeline of \product}

\product is rigorously and methodically tested across multiple stages within its CI environment: details have been reported by \pwork. To briefly summarise, the testing pipeline consists of four main phases: local testing, pre-submit testing, post-submit testing, and extended testing. During the development process, developers locally validate the new changes by creating new tests or reusing existing regression tests. Once new changes are submitted, they are once again tested in the pre-submit testing stage, before being integrated into their respective components and, ultimately, the main branch. A change is incorporated into the main branch if and only if it passes the pre-submit testing. After being merged into the main branch, changes go through post-submit testing, which is conducted daily using additional tests that require more resources: this further ensures that the current version of the software functions properly. Finally, once the main codebase becomes ready to be released, extended testing is conducted, using both automated and manual testing, to make sure that the new release candidate satisfies all requirements and lacks any regression.

We note that, in \product, tests are executed in the form of \textbf{test suites}, each of which includes multiple \textbf{test cases}.\footnote{Due to its large size and complexity, \product contains multiple test suites, each of which validates a certain functionality.} Test cases are modular and can share helper functions. The automated testing involves running a set of selected test suites in parallel.

\subsection{Flaky Test Problem in \product}
\label{sec:background:flaky}
\tikzstyle{fail-run} = [rectangle, minimum width=0.7cm, minimum height=0.5cm, text centered, draw=black, fill=red!50, rounded corners]
\tikzstyle{pass-run} = [rectangle, minimum width=0.7cm, minimum height=0.5cm, text centered, draw=black, fill=green!50, rounded corners]
\tikzstyle{arrow} = [->,>=stealth]

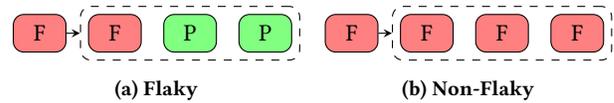
\begin{figure}[t]
  \centering
  \begin{subfigure}{.48\columnwidth}
    \centering
    \begin{tikzpicture}[main/.style = {draw}]
      \node (f-1) [fail-run] {F};
      \node (f-2) [fail-run, right of=f-1] {F};
      \node (f-3) [pass-run, right of=f-2] {P};
      \node (f-4) [pass-run, right of=f-3] {P};
      \node (all) [rectangle, draw=black, fit=(f-2) (f-3) (f-4), inner sep=1mm, rounded corners, dashed]  {};
      \draw [arrow] (f-1.east) -- (all.west);
    \end{tikzpicture}
    \caption{Flaky}
  \end{subfigure}
  \begin{subfigure}{.48\columnwidth}
    \centering
    \begin{tikzpicture}[main/.style = {draw}]
      \node (n-1) [fail-run] {F};
      \node (n-2) [fail-run, right of=n-1] {F};
      \node (n-3) [fail-run, right of=n-2] {F};
      \node (n-4) [fail-run, right of=n-3] {F};
      \node (all) [rectangle, draw=black, fit=(n-2) (n-3) (n-4), inner sep=1mm, rounded corners, dashed]  {};
      \draw [arrow] (n-1.east) -- (all.west);
    \end{tikzpicture}
    \caption{Non-Flaky}
  \end{subfigure}
  \caption{Example of flaky (left) and non-flaky (right) test results. Red and green rectangles represent failed and passed executions, respectively. After the first failure, each test is re-executed three times, which is denoted by the dashed line.}
  \label{fig:flaky-tests}
\end{figure}

We consider a test result flaky when the test both passes and fails in multiple executions against the same version of the program~\cite{parry_survey_2021}. Flaky test results are frequently observed during the testing of \product~\cite{Bach2022}. To identify flakiness in automated testing, \product adopts the popular \emph{rerun} strategy: if a test suite reports a failure, it is rerun typically three times to determine whether the failure is flaky or not. The process is depicted in Figure~\ref{fig:flaky-tests}.

This work specifically focuses on addressing the issue of flaky tests in the pre-submit testing phase of \product. Although flaky test results can be detrimental to any testing stage, they are particularly harmful to the pre-submit testing due to its higher frequency: according to \pwork, the pre-submit testing is performed about 80 times a day for the main branch, whereas the post-submit testing is done on a daily basis. Specifically, the flaky test issue induces the following two main problems in the pre-submit testing.

First, since pre-submit testing is performed more frequently than the subsequent phases, the rerun strategy consumes a much higher amount of machine resources when applied to the pre-submit testing. Our analysis of the historical CI data reveals that, often, hundreds of flaky test outcomes are produced per day. Consequently, multiple executions required by the rerun strategy not only incur a significant computational cost but also increase the overall turnaround time of testing, harming developer productivity. Our analysis of the past testing history of \product shows that on average 10\% of total testing time is spent on reruns to diagnose flakiness, with a maximum of 67\% when there are a large number of failures. A lightweight yet accurate technique that can predict whether an observed failure is flaky or not can reduce this cost.

Second, the large number of flaky results produced during the pre-submit testing phase also means that analysing and improving test flakiness would require a significant amount of human effort. Given its complexity, \product contains many potential causes of flakiness, such as bugs in source or test code, infrastructure issues, or external factors like errors in third-party libraries. The overwhelming number of flaky failures can lead developers to ignore flaky tests instead of analysing and improving them, potentially resulting in lower software quality standards. Such a loss of trust in the outcomes of tests can have a negative impact on the overall quality of the product~\cite{Rahman2018}. An automated analysis technique that groups flaky results according to their shared root causes can help developers deal with flaky tests more effectively.

To sum up, we aim not only to predict whether the first observed failure is flaky or not, but also to precisely group flaky results that share the same root cause, in the pre-submit testing phase.


\begin{figure}[t]
\begin{subfigure}{\columnwidth}
\centering
\begin{lstlisting}
$$/* stack trace #1, #2 */
Traceback (most recent call last):
  File ZZZ/ZZZ/NewDbTestCase.py line 937, in run
    self.setUp()
  File ZZZ/ZZZ/testCrossDBAtrMultiDB.py line 303, in setUp
    super(testCrossDBAtrMultiDB, self).setUp()
  File ZZZ/ZZZ/testCrossDBQuery.py line 1359, in setUp
    self.conn2 = self.conman2.createConnection()
  File ZZZ/ZZZ/connectionManager.py line 629, in createConnection
    return self.createNamedConnection(conn_id, **kw_args)
  File ZZZ/ZZZ/connectionManager.py line 704, in createNamedConnection
    **props)
  File ZZZ/ZZZ/connectionManager.py line 113, in __init__
    retryChecker(dbapi.Connection.__init__, self, **keys)
  File ZZZ/ZZZ/RetryChecker.py line 20, in __call__
    return function(*args, **kwargs)
$$/* error message #1 */
Error: (-10709, Connection failed (RTE:[89006] System call 'connect' failed, rc=111:Connection refused {~1.2.3.3:30024~ -> ~1.2.3.3:31144~} (~1.2.3.3:30024~ -> ~1.2.3.3:31144~)))
$$/* error message #2 */
Error: (-10709, "Connection failed (RTE:[89006] System call 'connect' failed, rc=111:Connection refused {~1.2.3.4:29616~ -> ~1.2.3.4:31144~} (~1.2.3.4:29616~ -> ~1.2.3.4:31144~))")
\end{lstlisting}
\caption{Symptoms of flaky failures}
\label{fig:failure-symptoms-flaky}
\end{subfigure}
\par\bigskip
\begin{subfigure}{\columnwidth}
\centering
\begin{lstlisting}
///* stack trace */
Traceback (most recent call last):
  File ZZZ/ZZZ/NewDbTestCase.py line 952, in run
    testMethod() # actually run the test
  File ZZZ/ZZZ/testCrossDBAtrMultiDB.py line 12487, in testCrossDB_ATR_BinaryDataSync_SubTable
    self._execute(cursors[1], """ALTER REMOTE SUBSCRIPTION "%s"."SUB_%s" DISTRIBUTE """ % (schemas[1], tables[1]))
  File ZZZ/ZZZ/testCrossDBAtrMultiDB.py line 429, in _execute
    self.fail("%s failed with %s" % (statement, str(err)))
///* error message */
AssertionError: ALTER REMOTE SUBSCRIPTION "db2"."SUB_tbl2" DISTRIBUTE  failed with (129, 'transaction rolled back by an internal error: table REP::db2:TARGET_tbl2 (t 2030) not locked by tablelock(false) or rowlock(false); $condition$=xlocked || rowlocked')
\end{lstlisting}
\caption{Symptoms of non-flaky failures}
\label{fig:failure-symptoms-non-flaky}
\end{subfigure}
\caption{Stack traces and error messages from the failures of the test case in \product.}
\Description{Examples of failure symptoms from \product}
\label{fig:failure-symptoms}
\end{figure}

\subsection{Motivating Example}
\label{sec:motivating-example}

Figure~\ref{fig:failure-symptoms} provides a detailed example to demonstrate the motivation behind our proposed approach. We have selected a test case in \product that has exhibited both flaky and non-flaky behaviour in the past. Figure~\ref{fig:failure-symptoms-flaky} shows parts of Python stack traces and error messages observed from two flaky failures of the test case. We can observe that the root cause of this flakiness is related to a database connectivity issue, as well as the specific call sequences that triggered this issue. Furthermore, the symptoms of flaky failures in this test case are distinguishable from those of non-flaky failures of the same test case, which are presented in Figure~\ref{fig:failure-symptoms-non-flaky}.
Based on this example, we conjecture that flaky failures sharing a root cause will result in similar error messages or stack traces, that are distinct from those of non-flaky failures. We also observe that it is common for flaky failures to be recurring across different pre-submit testing runs, as the underlying cause of the failure may not have been fully identified or resolved: in our example, the test case failed in 54 pre-submit testing runs between January and June in 2022. Out of these 54 failures, 52 (96\%) were caused by the same database connectivity issue, and their symptoms were exactly identical to those shown in Figure~\ref{fig:failure-symptoms-flaky}, except for the IP addresses highlighted in the grey background colour. These observations motivate us to detect recurring flaky failures using their symptoms.

\section{Just-in-Time Flakiness Detection Using Abstracted Failure Symptoms}
\label{sec:method}

This section presents our approach to detect flaky test failures during the pre-submit testing of \product using failure symptoms, e.g., stack traces and error messages. These are lightweight and black-box information sources that can be accessed without incurring additional execution or instrumentation costs. This allows us to design an efficient flakiness detection technique that can be seamlessly integrated into the CI pipeline in a just-in-time manner. The remainder of this section explains the details of our approach.

\tikzstyle{startstop} = [rectangle, rounded corners, minimum width=2cm, minimum height=1cm,text centered, draw=black, fill=white]
\tikzstyle{io} = [trapezium, trapezium left angle=70, trapezium right angle=110, minimum width=2cm, minimum height=1cm, text centered, draw=black, trapezium stretches=true, line width=0.5mm]
\tikzstyle{process} = [rectangle, minimum width=3cm, minimum height=1cm, text centered, draw=black, fill=white]
\tikzstyle{decision} = [diamond, minimum width=2cm, minimum height=1cm, text centered, fill=gray!20, draw=black, aspect=4]
\tikzstyle{arrow} = [thick,->,>=stealth]
\tikzstyle{arrow2} = [dashed,->,>=stealth]
\tikzstyle{storage} = [cylinder, draw=red!70!black,, text = black,
cylinder uses custom fill, cylinder body fill = red!10, cylinder end fill = red!40, aspect = 0.2, shape border rotate = 90]

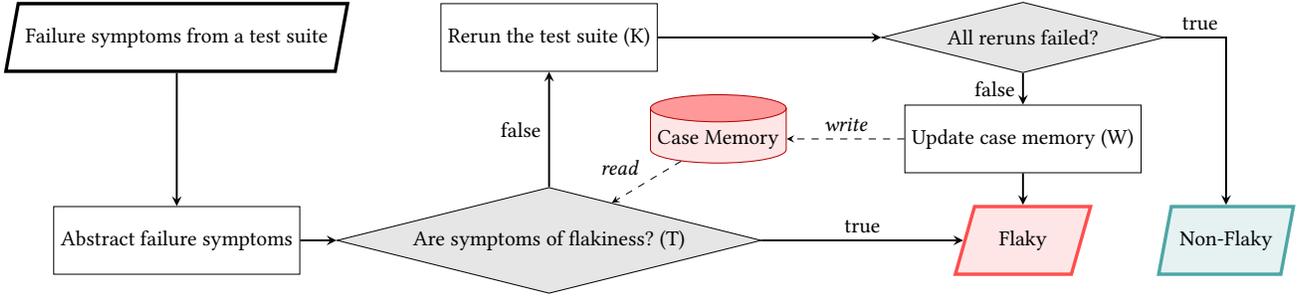
\begin{figure*}[ht]
  \scalebox{0.9}{
  \begin{tikzpicture}[node distance=2cm, auto]
    \node (in1) [io] {Failure symptoms from a test suite};
    \node (pro1) [process, below of=in1, yshift=-1cm] {Abstract failure symptoms};
    \node (dec1) [decision, right of=pro1, xshift=3.5cm, aspect=4] {Are symptoms of flakiness? (T)};
    \draw [arrow] (in1) -- (pro1);
    \draw [arrow] (pro1) -- (dec1);
    \node (out1) [io, right of=dec1, xshift=5cm, draw=red!70, fill = red!10] {Flaky};
    \draw [arrow] (dec1) -- node[anchor=south] {true} (out1);
    \node (pro3) [process, above of=dec1, yshift=1cm] {Rerun the test suite (K)};
    \draw [arrow] (dec1) -- node[anchor=east] {false} (pro3);
    \node (pro4) [process, above of=out1, yshift=-0.5cm] {Update case memory (W)};
    \node (dec2) [decision, above of=pro4, yshift=-0.5cm] {All reruns failed?};
    \draw [arrow] (pro3) -- (dec2);
    \draw [arrow] (dec2) -- node[anchor=east] {false} (pro4);
    \draw [arrow] (pro4) -- (out1);
    \node (out2) [io, right of=out1, xshift=1cm, draw=teal!70, fill = teal!10] {Non-Flaky};
    \draw [arrow] (dec2) -| node[anchor=south east] {true} (out2);
    \node (cm) [storage, left of = pro4, xshift=-2.5cm] {Case Memory};
    \draw [arrow2] (cm) -- node[anchor=south east] {\textit{read}} (dec1);
    \draw [arrow2] (pro4) -- node[anchor=south] {\textit{write}} (cm);
  \end{tikzpicture}}
  \caption{The overview of our flakiness detection approach during a pre-submit testing phase. The solid and dashed lines represent the control and data flow, respectively. The hyperparameters $T$, $K$, and $W$ respectively denote the minimum frequency threshold, the number of reruns, and the minimum word count for symptoms.}
  \Description{The overview of our flakiness detection approach}
  \label{fig:overview}
\end{figure*}

\subsection{Overview}
\label{sec:method:overview}
We propose a novel flakiness detection approach which is a hybrid of the conventional rerun strategy and the symptom-based flakiness detection. In our approach, the rerun strategy is used to systematically collect flaky test failures in a sound way, i.e., producing no false positives, whereas the symptoms of those failures are then regarded as a signal of flakiness and used to detect future flaky failures in a just-in-time manner.

The overall workflow of the proposed method is depicted in Figure~\ref{fig:overview}. Suppose that a test suite fails during the pre-submit testing phase. To decide whether to rerun the test suite, we first collect the set of failure symptoms, $S$, i.e., stack traces and error messages of the test suite failure. For example, if $N$ test cases within the test suite have failed, we collect the failure symptoms from each test case ($|S| = N$). Subsequently, we abstract each of the collected failure symptoms in $S$ by discarding less relevant details (Figure~\ref{fig:overview}: \textbf{Abstract failure symptoms}, see Section~\ref{sec:abstraction} for details). The set of abstracted symptoms, denoted as $S'$, are used to look up the case memory, $FFS$, which contains the known \textbf{F}laky \textbf{F}ailure \textbf{S}ymptoms. $FFS$ is a hash memory, where the key is the abstracted symptoms, and the value is the past observation count (default is $0$). To determine whether the currently observed symptoms $S'$ are the symptoms of flakiness or not, we use the following count-based matching function:
\[
    AreFlakinessSymptoms(S') :=
\begin{cases}
    true, & \text{if } \forall s \in S, FFS(s) \geq T\\
    false,& \text{otherwise}
\end{cases}
\]
If all collected symptoms in $S$ have a past observation count greater than a pre-defined threshold, $T$, the failure of the test suite is classified as flaky, and no further reruns are performed.
However, if at least one of their observation count is lower than $T$, we explicitly check the flakiness by rerunning it $K$ times (Figure~\ref{fig:overview}: \textbf{Rerun the test suite}).
If all $K$ reruns fail consistently, the failure is classified as non-flaky. Otherwise, the failure is classified as flaky, and the case memory is updated accordingly by incrementing the observation count of the symptoms of the failed test cases (Figure~\ref{fig:overview}: \textbf{Update case memory}). During this process, we heuristically filter out symptoms that are less likely to contain sufficient information about the root cause of the flakiness. We only store symptoms that have at least $W$ unique tokens with only alphabetic characters in their error messages. Let $S'_{W} \subseteq S'$ denote the set of symptoms that satisfy such a condition. Then, the case memory is updated for each of the symptoms in $S'_{W}$, i.e., $FFS(s) := FFS(s) + 1$ for all $s \in S'_{W}$.

Note that we maintain the case memory of flaky failures instead of the non-flaky ones for a specific reason. While the opposite approach, i.e., collecting and matching symptoms of non-flaky failures, is possible, it would be less accurate because of the inherent limitations of testing: a finite number of reruns can only prove flakiness, not non-flakiness. Consequently, symptoms of flaky failures can be collected reliably, while those of non-flaky failures cannot.

Our approach is matching-based~\cite{Brodie2005} rather than similarity-based~\cite{Rodrigues2022, Campbell2016, dang2012, bartz2008finding}, because matching is more efficient and scalable. Unlike hash-based matching with constant computational complexity, a similarity-based approach would require comparing the current symptoms with every past symptom, which is computationally expensive to be performed during testing. To further increase the effectiveness of matching, we abstract the failure symptoms to include only the information most relevant to the failure. The next subsection describes the details of the abstraction.

\begin{figure}[t]
  \centering
  \begin{subfigure}{\columnwidth}
  \begin{lstlisting}
  ZZZ/ZZZ/testCrossDBQuery.py,setUp
  ZZZ/ZZZ/connectionManager.py,createConnection
  ZZZ/ZZZ/connectionManager.py,createNamedConnection
  ZZZ/ZZZ/connectionManager.py,__init__
  ZZZ/ZZZ/RetryChecker.py,__call__
  \end{lstlisting}
  \caption{Example of the purified stack trace. After extracting only the file and function names from the original stack trace (Figure~\ref{fig:failure-symptoms-flaky}), the entry points for the test execution, i.e, the first two calls \texttt{run} and \texttt{setUp}, are discarded.}
  \Description{Example of the purified stack trace}
  \label{fig:purification}
  \end{subfigure}
  \par\bigskip
  \begin{subfigure}{\columnwidth}
  \begin{lstlisting}
  Error: (-~#~, Connection failed (RTE:[~#~] System call 'connect' failed, rc=~#~:Connection refused {~#.#.#.#:#~ -> ~#.#.#.#:#~} (~#.#.#.#:#~ -> ~#.#.#.#:#~)))
  \end{lstlisting}
  \caption{Example of the abstracted error message. The numbers in the original error message (Figure~\ref{fig:failure-symptoms-flaky}) are replaced with \texttt{\#} through the masking process.}
  \Description{Example of the abstracted error message}
  \label{fig:masking}
  \end{subfigure}
  \par\bigskip
  \begin{subfigure}{\columnwidth}
  \begin{lstlisting}
  [callstack]
  ZZZ/ZZZ/testCrossDBQuery.py,setUp
  ZZZ/ZZZ/connectionManager.py,createConnection
  ZZZ/ZZZ/connectionManager.py,createNamedConnection
  ZZZ/ZZZ/connectionManager.py,__init__
  ZZZ/ZZZ/RetryChecker.py,__call__
  [message]
  Error: (-#, Connection failed (RTE:[#] System call 'connect' failed, rc=#:Connection refused {#.#.#.#:# -> #.#.#.#:#} (#.#.#.#:# -> #.#.#.#:#)))
  \end{lstlisting}
  \caption{The abstracted stack trace and the error message are concatenated to form the symptom of a failure.}
  \Description{Example of the abstracted symptoms}
  \label{fig:concat}
  \end{subfigure}
  \caption{Example of the abstracted failure symptom}
  \Description{Example of the failure symptoms abstraction}
\end{figure}

\subsection{Abstraction of Failure Symptoms}
\label{sec:abstraction}

Symptoms of flaky failures with the same root cause may not be exact matches to each other, due to subtle differences such as the IP addresses in Figure~\ref{fig:failure-symptoms-flaky}. To achieve better matching, we propose to \emph{abstract} the failure symptoms to include only essential information related to the potential root causes, and to prevent minor differences from hindering correct matches. We apply purification and number masking to stack traces and error messages, respectively.

\noindent\textbf{Stack Trace Purification:} The test cases of \product are written as Python functions; consequently, most of the test failures are reported with their Python stack traces. The traces contain function names, line numbers, file names, and the source code line for each call frame unless the corresponding test suite is abnormally terminated. We purify the raw stack traces to contain only the essential information that captures the dynamic flow of the test execution: we first extract only the file and function names using regular expressions to filter out subtle differences in the stack trace, such as line number or source code style change. Subsequently, we remove entry points for test execution, i.e., the functions that are called to initiate the testing process, from the stack trace. This is to enable the matching of failure symptoms across different test suites that eventually trigger the same function sequences containing the root cause of flakiness. As a result, the stack trace is represented as a sequence of file and function pairs. Figure~\ref{fig:purification} shows the abstracted version of the raw stack trace from Figure~\ref{fig:failure-symptoms-flaky}.

\noindent\textbf{Number Masking:} We observe that many of the dynamic parts in the error message are numbers, e.g., IP addresses, dates, memory addresses, etc. To filter out such details, we replace all numbers in the error messages with the character \texttt{\#}. For example, the IP addresses in Figure~\ref{fig:failure-symptoms-flaky} are masked to \texttt{\#.\#.\#.\#:\#} as shown in Figure~\ref{fig:masking}. All hexadecimal numbers are also masked using the regular expression \texttt{0[xX][0-9a-fA-F]+}. This strategy is motivated by anonymization~\cite{amar2019mining}, abstraction~\cite{Nagappan2010}, removal of variables~\cite{He2017}, and removal of numbers~\cite{Salfner_Tschirpke_2008} in previous test log analysis techniques.

The abstracted stack traces and error messages are then concatenated to represent the symptom of a failure as shown in Figure~\ref{fig:concat}.

\section{Experimental Setup}
\label{sec:setup}
We describe the experimental setup to evaluate our approach.

\begin{table}[t]
  \caption{Statistics of 4,576 pre-submit testing records collected from \product. ``F'' represents flaky test failures, and ``NF'' represents non-flaky test failures.}
  \label{tab:dataset}
  \centering
  \scalebox{0.9}{
  \begin{tabular}{lr}
  \toprule
  \textbf{Statistics} & \textbf{Total}\\
  \midrule
  \# executed test suites & 8,750,036\\\midrule
  \# failed test suites (F+NF) & 58,927\\
  \# failed test suites (F) & 51,183\\
  \# failed test suites (NF) & 7,744\\\midrule
  \# failed test suites w/ test case failures (F+NF) & 15,114\\
  \# failed test suites w/ test case failures (F) & 11,599\\
  \# failed test suites w/ test case failures (NF) & 3,545\\\midrule
  \# failed test suites w/ valid symptoms (F+NF) & 13,168\\
  \# failed test suites w/ valid symptoms (F) & 9,857\\
  \# failed test suites w/ valid symptoms (NF) & 3,311\\
  \bottomrule
  \end{tabular}}
\end{table}

\subsection{Dataset Construction}
\label{sec:setup:dataset}
We evaluate our approach using the past pre-submit testing records from \product. Specifically, after collecting pre-submit testing results from January to June 2022, we assume that our technique was deployed in January 2022, with an empty corpus of flaky failure symptoms, and was used to detect the flakiness of future failures until June 2022. While \product has various combinations for the compiler and platform options for testing, we consider a single combination in our evaluation for the sake of simplicity. As a result, 4,576 pre-submit testing records are collected from the specified date range. Table~\ref{tab:dataset} shows the detailed statistics of the dataset, including the total number of executed test suites, the number of failed test suites (both flaky and non-flaky), the number of failed test suites with test case failures (both flaky and non-flaky), and the number of failed test suites with valid symptoms (both flaky and non-flaky). We observe that test suites often fail outside their test cases. For example, a test suite can crash during its setup or teardown process performed before and after the actual test case execution, or can be terminated due to timeout constraints. In such cases, the test suite cannot be associated with any test case failures. The data retention policy of \product does not keep the error messages in such cases for a long time, as such failures are outside the main testing processes. Therefore, we only consider the failures that occur during the execution of test cases in our analysis.\footnote{We note that our method can be extended to failures outside of test cases as long as the failure symptoms can be collected. This is discussed further in Section~\ref{sec:discussion}.} Furthermore, we found that some failure symptoms are not informative to be used as a signal for flakiness: for example, an error message ``Unit test failed - Log Preview not supported.'' does not provide any helpful information about its root cause. Therefore, we have manually mined a set of non-useful patterns of error messages and filtered out failures in our dataset that match the mined patterns. In total, we collected 13,168 test suite failures that occurred during the execution of test cases and have valid symptoms for every test case, corresponding to 22.3\% of the total failures. Among them, 9,857 failures are flaky, while 3,311 failures are non-flaky. This flakiness label is assigned based on the previous three rerun results of the failures. It should be noted that the non-flaky label may not be accurate as the reruns are not complete, i.e., they may not detect all flaky failures.

\subsection{Hyperparameter Settings and Other Details}
In our approach, three hyperparameters, $T$, $W$, and $K$, can be adjusted to optimise performance. First, the matching threshold, $T$, determines whether a given set of failure symptoms is an indicator of a flaky failure. A higher value of $T$ would lead to a more conservative detection. During our evaluation, $T$ is set with values of $\{1, 2, 3, 4, 5, 6\}$.
Second, the minimum required number of unique words in error messages, $W$, is used to heuristically control the quality of the failure symptoms. Like $T$, a higher value of $W$ would be more conservative. During our evaluation, $W$ is
set with values of $\{1, 2, 3, 4, 5, 6\}$. Lastly, the hyperparameter $K$ is used to determine the number of times each failed test suite should be rerun, and set to $3$ to align with the established practice in \product.

Furthermore, we assume that the pre-submit testing runs are executed sequentially in order of their starting time, for the sake of simplicity.
In addition, in a single pre-submit testing run, the case memory, $FFS$, is updated all at once after all necessary reruns for any failed test suites have been completed, because the test suites are executed in parallel in \product. We assume a sequential order between pre-submit test runs to ensure that the case memory is fully updated after each run, before the subsequent run starts.

\section{Results}
\label{sec:results}
In this section, we present the findings from our evaluation. Section~\ref{sec:results:accuracy} reports the accuracy of our symptom-based flakiness detection approach. Section~\ref{sec:results:abstraction} studies the impact of abstraction on the effectiveness of flakiness detection. Section~\ref{sec:results:savings} quantitatively analyses the potential savings in test resources achievable through our approach, as compared to the traditional rerun strategy. Finally, Section~\ref{sec:results:fp} provides a more in-depth analysis of false positive cases and discusses their implications.

\begin{figure*}[t]
  \centering
  \includegraphics[width=0.90\textwidth]{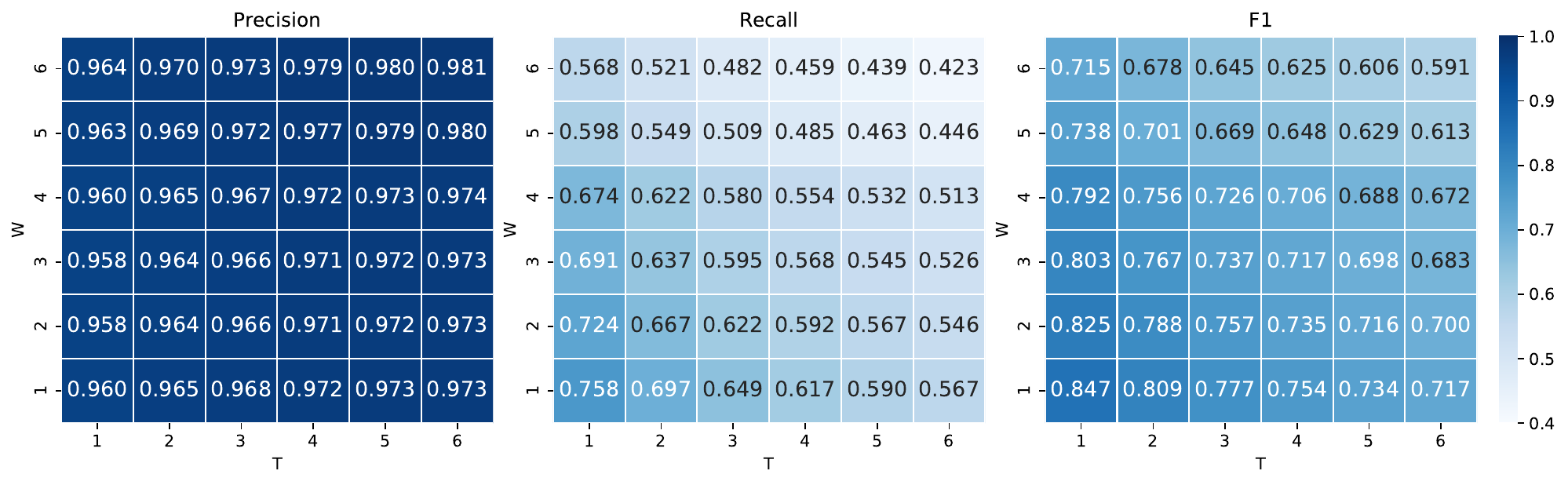}
  \caption{The performance of flaky failure detection on different hyperparameter settings. Each heatmap represents the precision, recall, and F1 of the detection results for each combination of $T$ and $W$. The darker the cell is, the higher the value is.}
  \Description{The performance of flaky failure detection on different hyperparameter settings for $T$ and $W$.}
  \label{fig:accuracy}
\end{figure*}

\subsection{Detection Accuracy}
\label{sec:results:accuracy}

We measure the overall accuracy of our flakiness detection approach based on the simulation results. For all failed test suites with valid failure symptoms, the precision and recall are defined as follows:
\[
  \text{Precision} = \frac{\text{\# actual flaky failures predicted as flaky}}{\text{\# failures predicted as flaky}}
\]
\[
  \text{Recall} = \frac{\text{\# actual flaky failures predicted as flaky}}{\text{\# actual flaky failures}}
\]
As a baseline for precision, we calculate the precision obtained from a naive model that always predicts positive, which is equivalent to the proportion of flaky examples in our dataset, $\frac{9,857}{13,168} \approx 0.749$.
To assess the overall performance of the detection, we also compute the F1 score, which is the harmonic mean of precision and recall.

The performance of our symptom-based detection approach is presented in 
Figure~\ref{fig:accuracy}, where we examine the precision, recall, and F1 
scores for different hyperparameter settings of $T$ and $W$.
Our approach consistently achieves precision of at least 
$0.958$, which is 28\% higher than the naive baseline. These results indicate 
that the failure symptoms can serve as effective indicators of flakiness.
In contrast, the recall values show a wider distribution, ranging from 0.423 to 
0.758, which in turn results in F1 scores ranging from 0.591 to 0.847. Overall, 
the best performing hyperparameter configuration is $T=1$, $W=1$, with the F1 
score of 0.847. Note that, due to the inherent nature of our method, recall 
values are sensitive to the frequency of recurring flaky failures. If a 
specific root cause of flaky failure manifests itself only once, our method 
will not be able to detect it, even under the least conservative hyperparameter 
configuration of $T=1$. As such, we note that the reported recall values are 
dependent on the specific data we used, i.e., the CI history from the six-month 
period.

We observe a trade-off between precision and recall against different 
hyperparameter settings. Increasing $T$ and $W$ leads to a more conservative 
detection approach, thereby increasing the precision, whereas lowering them 
would match more flaky symptoms and result in higher recall, saving more 
testing resources for reruns (See Section~\ref{sec:results:savings}). 
This trade-off provides the flexibility to finetune hyperparameters for meeting the specific requirements of the testing process. For example, if saving 
computational resources is the more pressing concern, one can opt for higher 
recall at the cost of spending human analysis cost to filter out false 
positives. On the other hand, if human analysis cost is the more pressing 
concern, one can finetune for higher precision and instead accept more reruns. 
In addition, other aspects can influence the selection of hyperparameters, such as whether there exists a subsequent safeguard (e.g., re-execution of 
all tests in the later post-submit testing stage) in the CI pipeline.


\subsection{Impact of Abstraction}
\label{sec:results:abstraction}

\begin{figure}
  \centering
  \includegraphics[width=\linewidth]{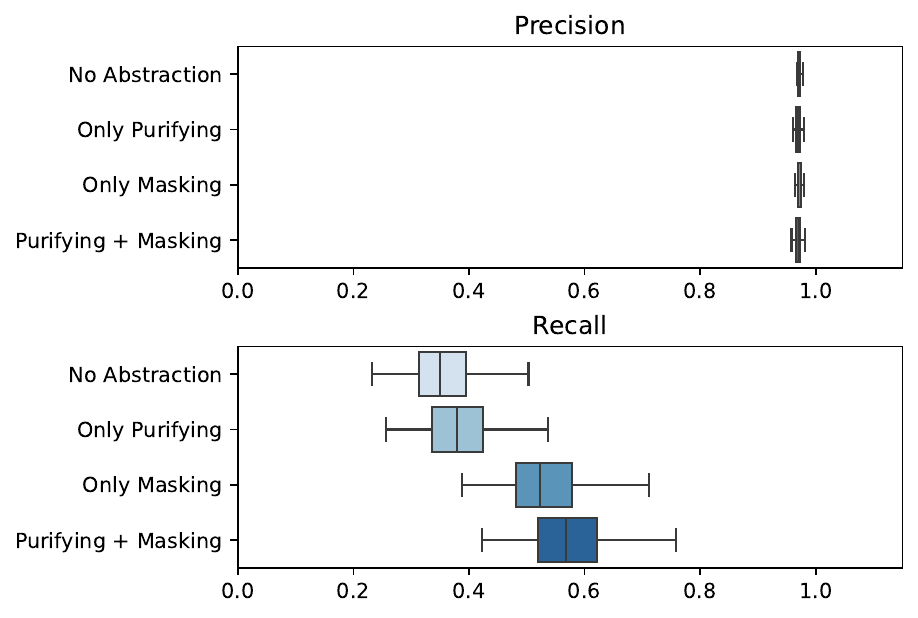}
  \caption{Precision (upper) and Recall (lower) for each abstraction setting: without abstraction, after only purifying stack trace, after only masking numbers, and after both purifying stack trace and masking numbers. The boxplots show the results for all hyperparameter settings.}
  \Description{Precision (upper) and Recall (lower) for each abstraction setting: without abstraction, after only purifying stack trace, after only masking numbers, and after both purifying stack trace and masking numbers. The boxplots show the results for all hyperparameter settings.}
  \label{fig:ablation}
\end{figure}
We perform an ablation study to see the impact of each of the abstraction methods on the performance of our flakiness detection approach. The boxplots in Figure~\ref{fig:ablation} show the precision and recall of our approach with different abstraction settings (y-axis). Note that each boxplot shows the precision and recall values across all tested hyperparameter settings.
Abstracting the failure symptoms increases recall on average by about 0.220 (from 0.352 to 0.572) against all hyperparameter values. At the best-performing hyperparameter configuration, $T=1$ and $W=1$, the abstraction increases the recall by 0.255 (from 0.503 to 0.758). While applying the symptom abstraction significantly increases recall, we can see that it does not sacrifice precision much; the precision drop is only 0.002 on average.
These results collectively show that abstraction enables more effective matching between flaky failure instances that have slightly different symptoms from each other.

\begin{figure}
  \centering
  \includegraphics[width=\linewidth]{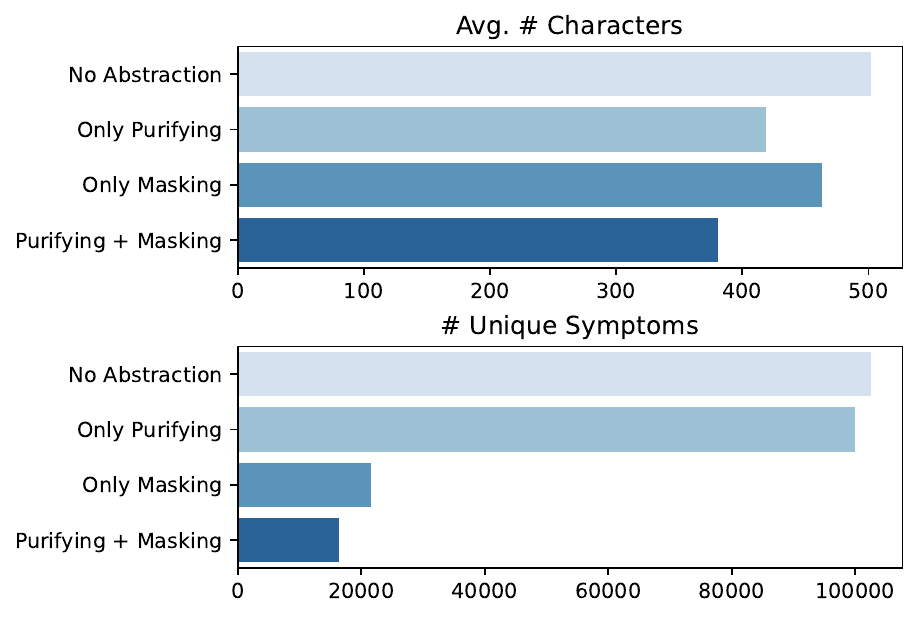}
  \caption{The average character length of symptoms (upper) and the number of unique failure symptoms (lower) for each abstraction stage: without abstraction, after only purifying stack trace, after only masking numbers, and after both purifying stack trace and masking numbers}
  \Description{The number of unique failure symptoms for each level of the symptoms abstraction.}
  \label{fig:unique}
\end{figure}

We also evaluate the direct impact of the abstraction on the symptoms. 
Figure~\ref{fig:unique} shows the average number of characters in failure 
symptoms (i.e., length) and the number of unique failure symptoms of test cases 
at each abstraction setting. The average length of failure symptoms decreases 
via abstraction, which is expected because the abstraction removes unnecessary 
information from the symptoms. We observe that the decrease in the number of 
unique failure symptoms is much more significant than the decrease in the 
length of symptoms: the number of unique failure symptoms is significantly 
reduced by the abstraction, from 102,529 to 16,345 (-84\%). Especially, masking 
numbers in error messages is effective in reducing the number of unique failure 
symptoms. This demonstrates that the abstraction of symptoms not only enhances 
the recall of our approach but also facilitates the grouping of similar 
failures into a single class based on the symptoms. For example, the abstracted 
symptoms in Figure~\ref{fig:concat} are matched to raw symptoms from 1,522 
failures across 120 test cases in 56 pre-submit testing runs. We expect that 
this automated failure grouping can help developers identify the root cause 
of flaky failures more efficiently.


\subsection{Resource Savings}
\label{sec:results:savings}
\begin{figure*}
  \centering
  \begin{subfigure}[b]{0.45\textwidth}
    \centering
    \includegraphics[width=\textwidth]{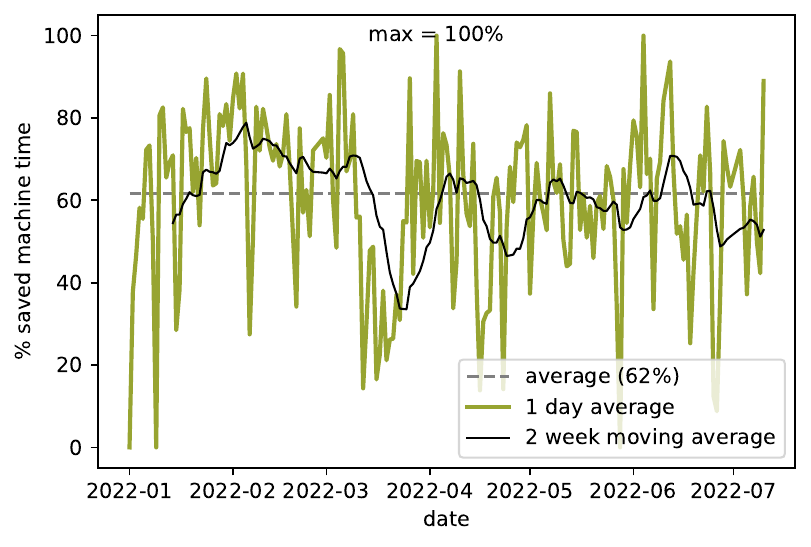}
    \caption{The reduction ratio of machine time}
    \Description{The reduction ratio of machine time}
    \label{fig:saved_time}
  \end{subfigure}
  \begin{subfigure}[b]{0.45\textwidth}
    \centering
    \includegraphics[width=\textwidth]{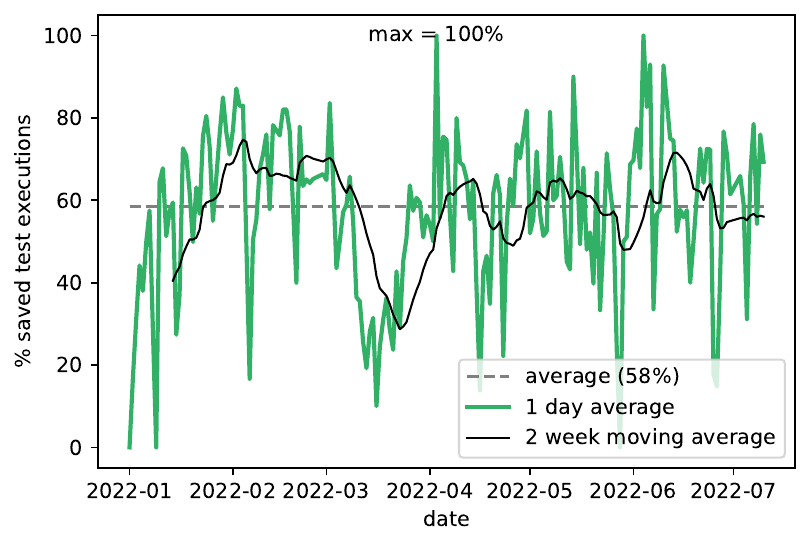}
    \caption{The reduction ratio of the number of test executions}
    \label{fig:saved_runs}
    \Description{The reduction ratio of the number of test executions}
  \end{subfigure}

  \caption{Percentages of the potential savings in the machine time and the number of test executions by our approach (with $T=1$ and $W=1$) compared to the rerun strategy for each day}
  \Description{Percentages of the machine time and the number of test executions saved by our approach compared to the rerun strategy}
\end{figure*}

\begin{figure}
  \centering
  \includegraphics[width=0.47\textwidth]{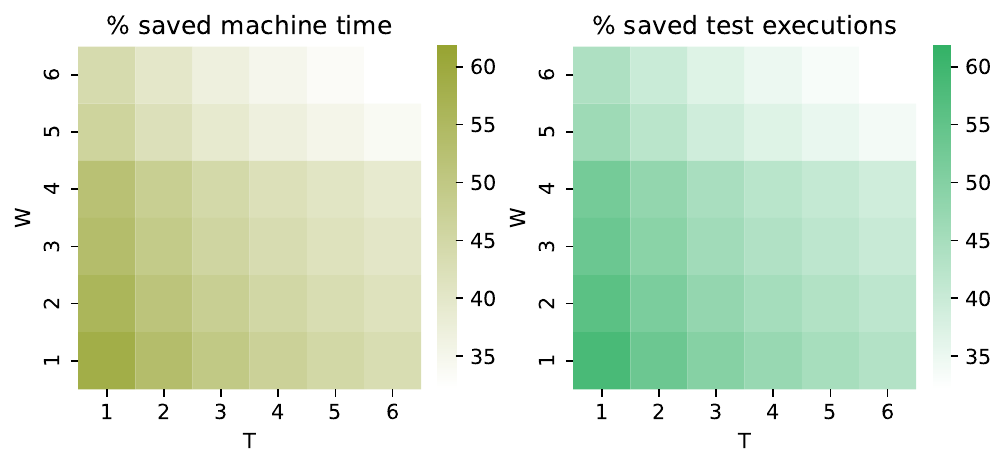}
  \caption{Heatmaps showing the average percentages of the machine time and the number of test executions saved by using our flakiness detection approach at each hyperparameter setting compared to the conventional rerun strategy}
  \Description{The ratio of machine time and test executions saved by our approach compared to the rerun strategy for each hyperparameter setting}
  \label{fig:saving_heatmap}
\end{figure}

We compute the percentages of the machine time and the number of test executions that can be saved by using our flakiness detection approach, compared to the conventional rerun strategy. Originally, each pre-submit testing run requires, on average, nine additional test executions and 3.13 hours of machine time for rerunning the studied failures (i.e., the 13,168 test failures with valid symptoms), which amounts to 219 executions and 78 hours spent per day.

Figure~\ref{fig:saved_time} and Figure~\ref{fig:saved_runs} show the total average, one-day average, and two-week moving average of the percentages of the machine time and the number of test executions saved by our approach with $T=1$ and $W=1$. Our approach can save up to 100\% of both test executions and machine time associated with reruns a day. On average across the entire period of evaluation, 62\% of machine time and 58\% of test executions can be saved when compared to the rerun strategy.
Figure~\ref{fig:saving_heatmap} shows the trend in the resource savings for every hyperparameter setting. Higher $T$ and $W$ values lead to lower resource savings, as they lead to fewer flaky failures being detected (i.e., lower recall). At the most conservative hyperparameter setting, $T=6$ and $W=6$, test executions and machine time could be potentially saved by 33\% and 32\%, respectively.
The results show that, although reducing both the values of $T$ and $W$ may not always be ideal (due to more false positives), it can effectively increase the cost savings in testing, particularly if coupled with an additional safeguard in subsequent testing stages. In the case of 
\product, since our symptom-based matching targets only the pre-submit testing 
stage, subsequent post-submit testing will still consider tests that have been 
flagged to be flaky in the pre-submit stage. Consequently, any false positives 
can be rectified through reruns in the post-submit testing.



\subsection{Analysis of False Positives}
\label{sec:results:fp}

The previous results in Section~\ref{sec:results:accuracy} show that our approach achieves a high precision of above 96\%. However, there is still a small number of \emph{false positive} test suite failures that are labelled as non-flaky in the dataset but predicted as flaky by our approach. Theoretically, these false positives can be classified into two categories:

\begin{itemize}
  \item (Case 1) The \emph{Non-Flaky} label is incorrect: As discussed in Section~\ref{sec:setup:dataset}, test suites are labelled based on the reruns with $K=3$. However, a limited number of reruns can still result in an incorrect "non-flaky" label~\cite{alshammari_flakeflagger_2021}. Failures in this category are not \emph{real} false positives from our approach, but rather the result of the limitations of the rerun strategy.

  \item (Case 2) The \emph{Flaky} prediction is incorrect: In instances where
 symptoms stored in the case memory are not a valid indicator of flakiness, they may lead our approach to incorrectly predict non-flaky test suite failures as flaky. Samples in this category are \emph{real} false positives.
\end{itemize}

Based on this categorisation, we analysed 82 false positive predictions that 
are shared by all hyperparameter settings.
To categorise false positive results, we first consult the historical manual review of the pre-submit test results. We conjecture that, if a failure belongs to Case 2 and is actually non-flaky (called \emph{test breakages} in \product), the corresponding code change will not be merged into the main branch. However, we find that, for 39 out of 82 false positive failures (48\%), the corresponding code changes have been successfully merged into the main branch after developers manually reviewed the test results. This suggests that the flaky label for these 39 failures is likely to be incorrect, i.e., they actually belong to Case 1 (we hereby refer to them as C1 candidates).
Since the testing system of \product allows us to re-trigger past pre-submit testing runs whose corresponding code change has been successfully merged, we attempted to rerun the C1 candidates to actually verify whether they are incorrectly labelled. Among the 39 C1 candidates, we were unable to verify 15 candidates due to limitations in the testing infrastructure or technical issues. The additional reruns for the remaining 24 C1 candidates reveal that all 24 are indeed verified to be flaky, i.e., their non-flaky labels are incorrect.
The analysis of C1 candidates suggests that at least 29\% (=24/82) of the initial false positive samples are actually true positives (i.e., Case 1) so that the actual precision and recall of our approach are higher than those reported in Section~\ref{sec:results:accuracy}.

\begin{figure}[t]
  \centering
  \begin{lstlisting}
  $$/* example 1 */
  failureException: one or more single tests failed in <UnitTests>
  $$/* example 2 */
  Test is marked as failed because it generated 1 unexpected output
  $$/* example 3 */
  Test failed. Most likely, the test crashed or hit the test timeout.
  There could be other reasons as well, e.g., subsequent XML update could not be written.
  $$/* example 4 */
  AssertionError: Test returned errorcode (rc = 2), error:
  $$/* example 5 */
  Failure in 1 out of 32 batched results. Failed tests: _traceMdsSc3.QMDS5781Queries in batch: test_traceMdsSc3.QMDS5755,|<...stripped...>|,test_traceMdsSc3.QMDS5787
  \end{lstlisting}
  \caption{Examples of uninformative error messages that lead to incorrect flaky failure detection}
  \Description{Examples of uninformative error messages that lead to incorrect flaky failure detection}
  \label{fig:fp}
\end{figure}

To determine the cause of incorrect predictions for Case 2, we have manually examined the failure symptoms of the remaining 43 false positive predictions. By definition, false positives in Case 2 mean that some symptoms stored in, and matched from, the case memory are not exclusive to flaky failures. We find that most of these symptoms are uninformative and vague, despite our attempt to filter out such symptoms (see Section~\ref{sec:setup:dataset}). Figure~\ref{fig:fp} shows false positive symptoms from Case 2: they only indicate that a failure has occurred, without providing any information on the internal program states or the location where the crash occurred.
This shows that, in order to further enhance the precision of our approach, it is necessary to either construct a more thorough list of patterns of uninformative symptoms to filter them out or, more fundamentally, improve the quality of the test cases so that their failure symptoms contain more meaningful information.
In this regard, a careful manual analysis of false positives from the historical data (i.e., non-flaky test runs confirmed by reruns) can be useful not only for improving the precision of the proposed technique (by filtering out unhelpful symptoms), but also for improving the overall quality of tests (by rewriting them to be more informative).

\section{Discussion}
\label{sec:discussion}

This section presents the developer feedback on the usefulness of failure symptoms and discusses about a potential extension of our approach to detect other flaky failures in \product.

\subsection{Developer Feedback on the Usefulness of Failure Symptoms in Debugging}

We collected a subset of abstracted failure symptoms for flaky failures whose observation counts are more than 20. We asked developers of \product for their assessments of the usefulness of these symptoms in identifying the root causes of the flakiness. The objective of this exercise is to gain an understanding of the developers' perspectives on the utility of the failure symptoms in the debugging process. Note that the answers and feedback can be biased by the experience of the developers.

The feedback from developers about the abstracted failure symptoms collected so far is mixed. Some of the failure symptoms are considered to be valuable in determining the source of the flakiness. For example, symptoms that display specific types of errors (such as timeouts, missing attribute errors, or import issues), or those that include specific file names or program components known to cause flakiness, are considered to be effective indicators of potential root causes. However, some more general symptoms are seen as less helpful, and may not provide enough information to pinpoint the source of the problem. These symptoms are often too generic and lack specificity. For instance, some symptoms consist only of a stack trace with generic file and method names frequently used by many test cases, or an error message that is too brief, e.g., \texttt{AssertionError: \# != \#}. Essentially, these are symptoms that are similar to Case 2 false positives, described in Section~\ref{sec:results:fp}. They do not highlight the cause of flakiness and can be produced by failures due to a variety of reasons, including the test suite, the testing environment, or the program being tested.

The feedback collectively highlights the importance of having informative failure symptoms for effectively detecting and debugging flaky tests. Without detailed symptoms, it becomes challenging or even impossible for developers to accurately determine the source of the failure. In turn, this points to the importance of writing test cases of \product with more descriptive error messages that clearly indicate the issue including information about relevant program states and any other details that can aid in diagnosis. 



\subsection{Addressing Failures Outside of the Main Testing Body}

Let us consider the test failures that occur outside the scope of our evaluation. As explained in Section~\ref{sec:background:flaky}, the test suite of \product is composed of multiple test cases, which form the main testing body that validates the behaviour of the program. To run the test cases, the test suite first sets up the testing environment, executes the test cases, and then tears down the environment.
In Section~\ref{sec:setup:dataset}, we observe that a large number of flaky failures in \product happen outside of the main testing body. Only 15,114 (or 30\%) of the failures occur during the execution of test cases. We have sampled and manually investigated some of the remaining 70\% of failures, and identified two primary reasons for these failures: (1) the setup/teardown part outside of the test cases leads to exceptions or errors (including a timeout) (2) a timeout occurs while executing the test cases, but is not handled gracefully, leaving no meaningful symptom. These points are in line with the results from a previous study~\cite{parry_surveying_2022}, which found that developers rate issues with setup/teardown to be the most common causes of flakiness.

During the evaluation, we have focused on the flaky failures that occur in the main body of the test suites, because those are the only failures for which we can collect past symptom data. However, we argue that our approach can potentially be extended to address remaining flaky failures that occur during the setup and teardown process, as long as they produce valid failure symptoms. Under these circumstances, the set of failure symptoms for a given test suite, $S$, in Section~\ref{sec:method:overview} can be a singleton set containing a symptom from the setup/teardown failure. Similarly, test suites that time out without leaving any meaningful symptoms should be improved with better graceful shutdown mechanisms, so that they can produce more informative error messages that summarise the timeout context. We believe that these suggestions, along with the recommendations mentioned in the previous section, can serve as a guide for improving the failure handling practices in \product.

\section{Related Work}
\label{sec:rw}

This section covers the related work on flaky test detection and failure deduplication.

\subsection{Flaky Test Detection}
\label{sec:rw:flaky}

A widely adopted way to detect flaky tests is the rerun strategy~\cite{google_flaky,spotify_flaky}, i.e., to rerun the failed test cases multiple times and check if they eventually pass or not. Gruber et al.~\cite{gruber_empirical_2021} find that a large number of reruns is needed to diagnose test flakiness. However, doing numerous reruns is not feasible in practice due to its high cost. To address this challenge, several techniques have been proposed to detect flaky tests without rerunning them.

First, there is a group of techniques that use dynamic features of test executions to detect flakiness. DeFlaker~\cite{bell_deflaker_2018} uses coverage information to detect flaky tests that do not execute any of the changed code. However, the size of \product and the overhead for coverage collection forced us to collect coverage only on a weekly basis instead of for every pre-submit testing run. Consequently, we cannot explicitly check whether failing tests cover any of the recently changed code, making it difficult to apply DeFlaker to our use case.
FlakeFlagger~\cite{alshammari_flakeflagger_2021} trains a machine learning model that takes both static code features including test smells and dynamic features such as coverage as input, and predicts whether a given test failure is flaky, with up to 86\% accuracy. Among the studied features, dynamic information such as execution time and coverage is found to be the most important features.

Second, some approaches look at previous test execution histories to detect flaky failures. Herzig et al.~\cite{herzig_empirically_2015} collect both test features (e.g., test case identifier) and test results (i.e., passed or failed), and subsequently use association rule learning to identify patterns of the flaky test results. Kowalczyk et al.~\cite{kowalczyk_modeling_2020} quantitatively model the flakiness of a test case based on the temporal variance of its results during test history. Gruber et al.~\cite{gruber2023practical} uses several features related to the code evolution and test history data, such as the number of changed files in the most recent pull request or the flip rates of test outcomes, to detect flakiness without reruns.


Last, there are approaches that use static features of the given program, such as source code tokens contained in the test code~\cite{pinto_what_2020, camara_what_2021, haben_replication_2021} or test smells~\cite{pontillo_toward_2021, pontillo_static_2022}, to detect the flaky tests.
Pinto et al.~\cite{pinto_what_2020} identify the vocabulary of flaky test cases, i.e., tokens such as job, action, and services that are highly associated with flakiness, and show that a model that solely depends on static code features can achieve high accuracy on flakiness detection.
Pontillo et al.~\cite{pontillo_toward_2021} investigate the difference between flaky and non-flaky tests in terms of 25 code metrics and smells, which has later been replicated~\cite{pontillo_static_2022} using a different dataset from another study~\cite{alshammari_flakeflagger_2021}.

Our approach can be considered a hybrid of all existing techniques. We use 
dynamic features from test executions, but only those that do not require 
costly code instrumentation such as stack traces and error messages. We 
consider test execution history, but instead of focusing on the overall test 
outcome of pass or fail, we maintain a case memory focused on the symptoms of 
flaky failures. Finally, while some of the symptoms we collect are part of the 
source code, they are not static, as we obtain them via test executions. Note 
that the same test case can result in both flaky and non-flaky failures without 
changing its source code, as shown in Figure~\ref{fig:failure-symptoms}. Our 
approach can accommodate such variability because it collects failure symptoms 
dynamically: in contrast, a static approach will permanently label a test case 
as flaky or not as long as its source code does not change.

\subsection{Failure Deduplication}

Flakiness detection can be considered a specific form of failure root cause analysis. Jiang et al.~\cite{jiang_causes_2017} aim at identifying causes of test failures from predefined categories, including test flakiness. The suggested technique matches the current test log output with textually similar past logs. As a result, the technique can suggest detailed causes of the current test failure, which could be more informative than a binary flakiness label.

Another form of failure root cause analysis is failure deduplication, i.e., grouping test failures based on shared root causes~\cite{Brodie, bartz2008finding, joshy_2022, dang2012, Rodrigues2022,lerch2013finding}.
Since deduplication typically follows test execution, outputs of failures, such as stack traces and error messages, are often adopted by failure deduplication techniques as inputs: the intuition is that the more similar two stack traces or error messages are, the more likely that the corresponding failures share the common root cause.
Bartz et al.~\cite{bartz2008finding} confirmed that the edit distance between two stack traces is an important feature for a machine learning classifier trained to identify failures with shared root causes. Lerch et al.~\cite{lerch2013finding} reported that stack traces are the most valuable information contained in bug reports that can be used for deduplication.

Since not all failures caused by the same root cause exhibit the exactly same stack trace, various ways of abstracting stack traces have been suggested. Brodie et al.~\cite{Brodie} filter out entry points and common error handling routines using a ``stop-words'' list provided by a domain expert, and remove recursive function calls, before matching stack traces.
Modani et al.~\cite{Modani2007} also remove less relevant functions from stack traces before measuring either the edit distance or the length of the longest common subsequence between them. Joshy et al.~\cite{joshy_2022} use only the top $N$ calls on the stack trace to group the failures. Our stack trace purification is similarly motivated.

There are more complicated approaches for comparing two stack traces to group similar failures. Dang et al.~\cite{dang2012} proposed a weighted common subsequence measure to quantify the similarity between two stack traces. Rodrigues et al.~\cite{Rodrigues2022} designed a new similarity metric between two stack traces based on the optimal global alignment between them. However, as similarity calculation is computationally expensive, these approaches are hard to be applied to our just-in-time flakiness detection.

In addition to the stack traces, error messages, which contain unstructured information about exception types and output values, have been also used to group similar failures. Erman et al.~\cite{erman2015navigating} used raw error messages, together with the test case names, to cluster test results. CloudBuild~\cite{Esfahani2016}, a Microsoft's build-and-test system, contains a flaky test management system called Flakes, which groups and reports flaky failures with similar error messages together~\cite{lam_study_2020}.
However, our approach differs from the similarity-based approaches as it uses hash-based matching. While matching is much more efficient, it is susceptible to irrelevant details. We apply the number masking to prevent such details from hindering exact matching between similar flaky failures; our masking method is inspired by prior studies on test log analysis~\cite{amar2019mining, Nagappan2010, Salfner_Tschirpke_2008}.

\section{Conclusion}
\label{sec:conclusion}

We report our experience of using failure symptoms as a means to detect flaky failures in \product in a just-in-time manner. Our approach is inspired by previous failure deduplication studies. We collect symptoms of flaky failure using the conventional rerun strategy, and later detect flaky failures by matching their symptoms to the previous flaky failures, with a precision of up to 98\%. Our empirical evaluation with real-world CI data from \product, along with feedback from developers, yields the following takeaways:
1) Stack traces and error messages are a valuable resource for recognising flaky test failures in a CI pipeline;
2) Abstracting failure symptoms can significantly increase the recall of flakiness prediction, while allowing the automated grouping of flaky failures;
3) Having tests produce detailed and informative failure symptoms is crucial to the accurate detection and debugging of flaky tests.
In future work, we aim to reduce the false positive rate of our approach by automatically detecting and filtering out uninformative symptoms. In the longer term, we hope that our analysis of the failure symptoms can guide the developers to improve the quality of test outputs.


\balance
\bibliographystyle{ACM-Reference-Format}
\bibliography{ref}

\end{document}